\documentclass[aps,floatfix,twocolumn,tightenlines,amsmath,amssymb]{revtex4-1}

\usepackage{graphicx}
\usepackage{amssymb}
\usepackage{color}
\usepackage{psfrag}
\usepackage{ifsym}
\usepackage{epstopdf}
\usepackage{dsfont}
\usepackage{amsmath}
\usepackage{multirow} 
\usepackage{paralist}
\usepackage{xfrac}

\usepackage{amsthm}

\newcommand{\ket}[1] {| #1 \rangle}
\newcommand{\braket}[2] {\langle #1 | #2 \rangle}
\newcommand{\ketbra}[1] {| #1 \rangle\!\langle #1 |}

\newcommand {\be} {\begin{eqnarray*}}
\newcommand {\ee} {\end{eqnarray*}}
\newcommand {\bea} {\begin{eqnarray}}
\newcommand {\eea} {\end{eqnarray}}

\begin{document}
\title{Information Causality in the Quantum and Post-Quantum Regime}

\author{Martin Ringbauer$^{1,2}$\footnote{Electronic address: {ringbauer.martin@gmail.com}}, Alessandro Fedrizzi$^{1,2}$, Dominic W.\ Berry$^{3}$ and Andrew G. White$^{1,2}$}
\affiliation{$^1$Centre for Engineered Quantum Systems, $^{2}$Centre for Quantum Computation and Communication Technology, School of Mathematics and Physics, University of Queensland, Brisbane,   QLD 4072, Australia.\\
$^3$Department of Physics and Astronomy, Macquarie University, Sydney, NSW 2109, Australia}

\begin{abstract}
Quantum correlations can be stronger than anything achieved by classical systems, yet they are not reaching the limit imposed by relativity. The principle of information causality offers a possible explanation for why the world is quantum and why there appear to be no even stronger correlations. Generalizing the no-signaling condition it suggests that the amount of accessible information must not be larger than the amount of transmitted information. Here we study this principle experimentally in the classical, quantum and post-quantum regimes. We simulate correlations that are stronger than allowed by quantum mechanics by exploiting the effect of polarization-dependent loss in a photonic Bell-test experiment. Our method also applies to other fundamental principles and our results highlight the special importance of anisotropic regions of the no-signalling polytope in the study of fundamental principles.
\end{abstract}

\maketitle

\section{Introduction}
Quantum mechanics is one in a large class of theories which are consistent with relativity in the sense that they do not allow signals to be sent faster than the speed of light. Many of these theories exhibit strong non-local correlations between distant particles that cannot be explained by the properties of the individual particles alone. Surprisingly, quantum mechanics is not the most non-local among them, which raises the question about the physical principle that singles out quantum mechanics and sets the limit on the possible strength of correlations in nature.

Here we experimentally address this fundamental question by testing the principle of information causality in the classical, quantum and post-quantum regime. While the no-signaling principle limits the speed with which distant parties can communicate, information causality states that the accessible information cannot be more than the information content of a communicated message, no matter what other shared resources are used. Both classical and quantum mechanics satisfy this principle, while it is violated by most post-quantum theories~\cite{Allcock2009}.

We experimentally emulate correlations of various strengths from classical to almost maximally non-local and demonstrate a violation of the principle of information causality in the case where the simulated correlations are beyond the quantum regime. Apparent super-quantum correlations are, in our approach, a consequence of the non-unitary evolution of quantum states when subjected to polarization-dependent loss with post-selection~\cite{Berry2010}. For moderate loss, we find that initially entangled states can result in super-quantum correlations, while unentangled states still appear classical. For higher loss on the other hand we observe super-quantum correlations even for classical input states.

\section{Theoretical framework}
No-signaling resources can formally be treated as pairs of black boxes shared between arbitrarily separated Alice and Bob~\cite{Barrett2005}, see Fig.~\ref{fig:ICMotivation}a). Each box has a single input and output and the correlation between them is only restricted by the no-signaling principle. This means that the local outcome only depends on the local input, such that Alice cannot learn anything about Bob's input from only her output.

A typical quantum example of such a resource is a pair of entangled particles, shared between Alice and Bob, where inputs correspond to measurement settings and outputs to measurement outcomes. Since the work of John Bell---and numerous subsequent confirming experiments---it is now widely accepted that these particles exhibit non-local correlations, which have no classical explanation. Under the no-signaling constraint alone, however, there are even stronger non-local correlations than quantum entanglement~\cite{Popescu1994}. The maximum that is compatible with relativity is achieved by the so-called Popescu-Rohrlich (PR)-box~\cite{Popescu1994}, characterized by perfect correlations of the form $A\oplus B = ab$, between Alice's and Bob's inputs $a$ and $b$ and outputs $A$ and $B$, respectively. Here $\oplus$ denotes addition modulo $2$, equivalent to the logical XOR, where $A\oplus B=0$ when $A=B$ and $1$ otherwise.

A convenient operational way of quantifying non-locality is the Clauser-Horne-Shimony-Holt (CHSH) inequality~\cite{Clauser1969}. This experimentally testable reformulation of Bell's inequality is satisfied by any correlation that can be described by a local hidden variable model. Such models are a description of correlations that can arise in classical systems, but cannot describe non-local correlations obtained from e.g.\ entangled quantum states. Written in terms of correlations of the form $A \oplus B=ab$ the inequality takes the form
\begin{equation}
S = \sum_{a=0}^1 \sum_{b=0}^1 P(A\oplus B = ab\mid a, b) \leq 3 .
\label{eq:CHSH}
\end{equation}
Here $P(A\oplus B = ab\mid a, b)$ denotes the probability for obtaining outputs $A,B$, which satisfy $A\oplus B=ab$ given the inputs $a$ for Alice and $b$ for Bob.
While this inequality is satisfied by any classical correlations, it can be violated in the quantum case. This violation, however, is bounded to a value of $2+\sqrt{2}\approx 3.41$, known as Tsirelson's bound~\cite{Cirel'son1980}. Note, that inequality~\eqref{eq:CHSH} is presented here in a slightly different form than conventionally~\cite{Clauser1969}, where the classical bound is $2$ and Tsirelson's bound is $2\sqrt{2}$. They are, however, linearly related and the difference is a simple rescaling of $S$.

Despite being a simple consequence of the mathematical formalism of quantum mechanics, it is unclear what the physical motivation is for this seemingly sub-optimal limit on the strength of quantum correlations. In fact even the algebraic maximum $S=4$ can be achieved (by the PR-box) without violating the no-signaling principle.

\section{The principle of information causality}
This principle is physically motivated by the fact that, according to special relativity, faster-than-light information transfer would allow information to be sent backwards in time and thus violate causality. Nevertheless, it does not explain why super-quantum correlations such as the PR-box are incompatible with quantum mechanics and seem not to exist in nature. A possible explanation is offered by the principle of information causality---a generalization of no-signaling---which states that there cannot be more information available than was transmitted~\cite{Pawlowski2009}.

This can be understood on the basis of the following elementary information-theoretic protocol: Bob tries to gain information from a set of data that is only known to Alice. The parties are allowed to use an arbitrary amount of shared no-signaling resources, but may not communicate more than $m$ classical bits. In this case, the information causality principle states that the amount of information accessible to Bob should be limited to $m$ classical bits~\cite{Pawlowski2009}.

\begin{figure}[h!]
\begin{center}
\includegraphics[width=1\columnwidth]{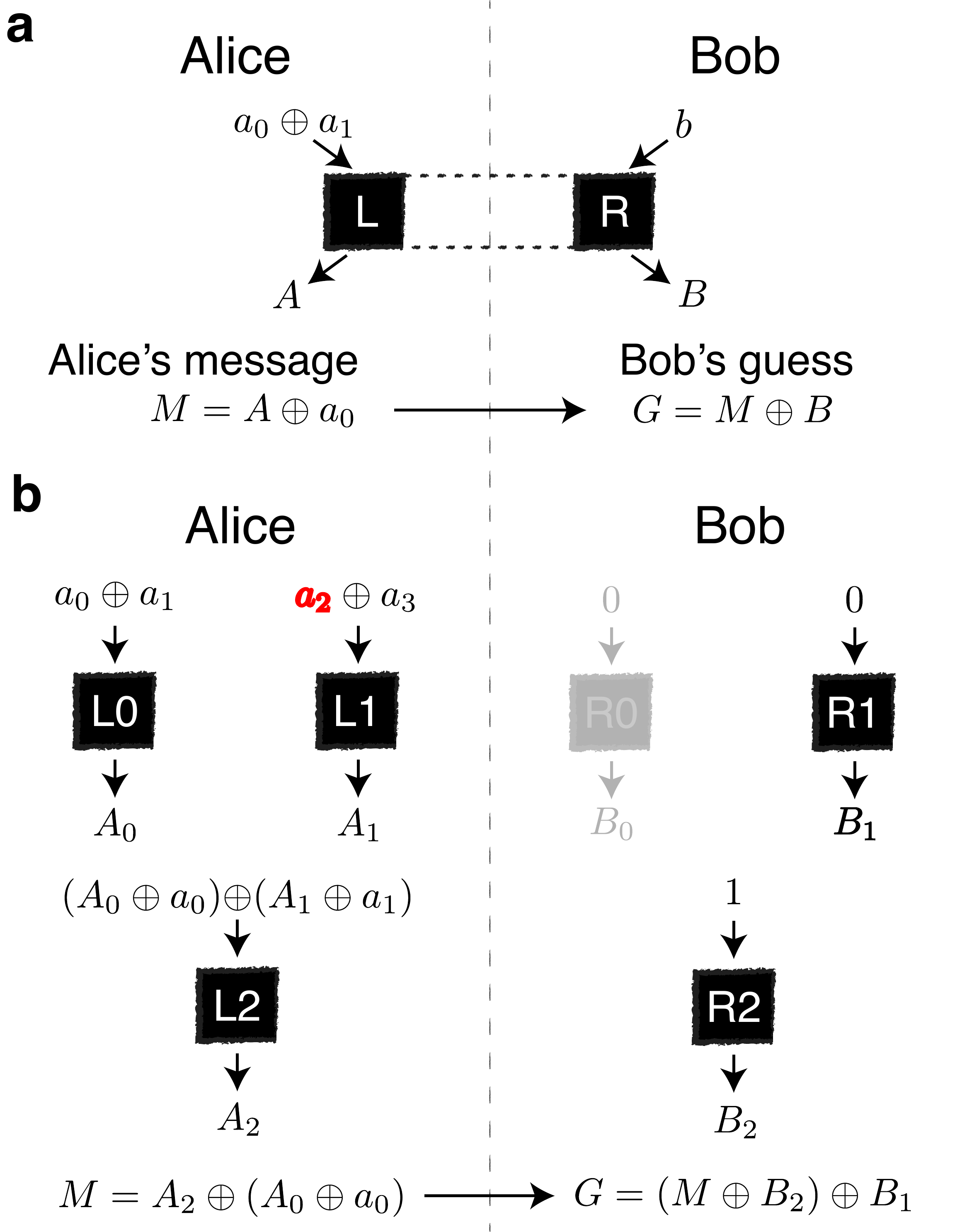}
\end{center}
\caption{\textbf{Illustration of the information causality protocol.} \textbf{a)} A general no-signaling resource is given by a space-like separated (indicated by the dashed line) pair of black boxes producing local outputs $A$ and $B$ for Alice and Bob, when they input $a$ and $b$, respectively. In the case of a PR-box the outputs of the left (L) and right (R) box would be perfectly correlated according to $A\oplus B = ab$. The inputs and outputs depicted here correspond to the simplest instance of the information causality protocol.
\textbf{b)} Example of the multilevel information causality protocol for $n=2$. Alice has a list of $N$ bits $a_i$ and Bob tries to guess the bit $a_3$ (shown in bold, red) using $N{-}1{=}3$ pairs of shared black boxes on $n{=}2$ levels (corresponding boxes labeled $L0/R0, L1/R1$, $L2/R2$). Bob's inputs $b_i$ and choice of boxes are determined by the binary decomposition $b=\sum_{k=0}^{n-1}b_k 2^k$. From his outputs $B_1,B_2$ and Alice's $1$-bit message $M$ Bob computes a final guess $G$ for Alice's bit $a_b$. Note that Bob only needs to use one box on each level and ignores the outputs of all the other boxes. Hence, his input to these boxes can be arbitrary and in the experiment we chose to use the same input for all boxes on one level.}
\label{fig:ICMotivation}
\end{figure}

In the simplest instance, Alice has a set of two bits $\{a_0,a_1\}$ and Bob wants to guess one of them, which we denote $a_b$~\cite{VanDam2000}, see Fig.~\ref{fig:ICMotivation}a). Alice and Bob then input $a_0 \oplus a_1$ and $b$ into their respective black box and obtain outputs $A$ and $B$. From this output Alice computes an $m=1$-bit message $M = A\oplus a_0$ and sends it to Bob, who calculates his guess for Alice's bit as $G = M \oplus B = a_0\oplus A\oplus B$. In the case of a shared PR-box, Bob can guess either one of Alice's bits perfectly, since in that case $A\oplus B = ab$ and thus $G = a_0 \oplus b(a_0 \oplus a_1)$.

In the more general case considered here, Alice has a dataset $\{a_0,\ldots,a_{N-1}\}$ of $N=2^n$ bits and Bob wants to guess the bit with index $b=\sum_{k=0}^{n-1}b_k 2^k$. As discussed in Ref.~\cite{Pawlowski2009}, Alice and Bob can achieve this task by using a nested version of the protocol outlined above, with $N-1$ black boxes on $n$ levels and $1$ bit of classical communication.

The protocol is illustrated in Fig.~\ref{fig:ICMotivation}b) for the case $n=2$. From every output Alice computes a temporary message $M_{k,i}$, where $k$ denotes the level and $i$ the number of the box on that level. Since she is only allowed $1$ bit of communication, she uses these temporary messages as the inputs for the boxes on the next-lower level and only sends the final message to Bob. Depending on $b_n$ Bob then decodes either $M_{n-1,1}$ or $M_{n-1,2}$ and then moves on to the next-higher level until he reaches the bit of interest.

Bob's success can then be quantified by
\begin{equation}
I = \sum_{k=0}^{N-1} I(a_k\colon G\mid b=k) ,
\label{eq:ICQuantity}
\end{equation}
where $I(a_k\colon G\mid b=k)$ is the Shannon mutual information between the $k$'th bit of Alice's list and Bob's guess for it~\cite{Pawlowski2009}. This quantity can further be bounded as
\begin{equation}
I \geq \sum_{k=0}^{N-1} 1- h(P_k) ,
\label{eq:ICbound}
\end{equation}
where $h(P_k)$ is the binary entropy of the success probability $P_k$ for guessing the $k$'th bit.

\section{Experimental implementation}
\begin{figure}[h!]
\begin{center}
\includegraphics[width=1\columnwidth]{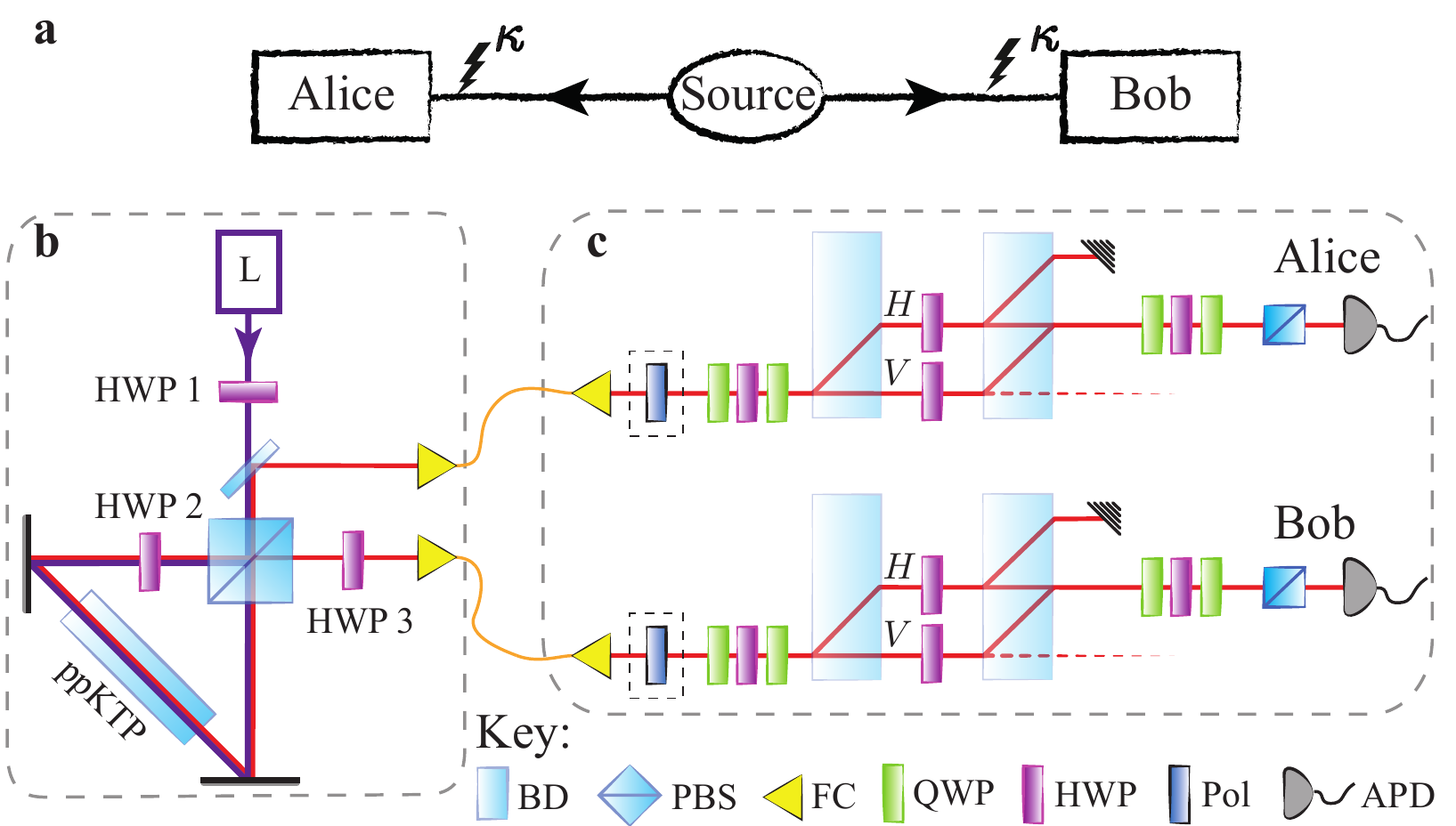}
\end{center}
\caption{\textbf{The experimental approach.} \textbf{a)} Pairs of single photons are created at the source and are subjected to polarization-dependent loss before Alice and Bob perform their measurements. \textbf{b)} The photon-source used in the experiment is spontaneous parametric down-conversion in a $10$~mm long periodically poled KTiOPO$_4$ (ppKTP) crystal inside a polarization Sagnac interferometer using a grating stabilized continuous wave pump laser (L) at a wavelength of $\lambda=410$~nm. By controlling the phase and polarization of this laser and adjusting the additional half-wave-plate in Bob's arm of the source, HWP3, any two-qubit states can be produced. 
\textbf{c)} Polarization-dependent loss is introduced to the system in a controllable way using an interferometer based on calcite beam displacers (BD), which split the horizontal and vertical polarization components into two spatial modes. The two HWPs in the interferometer are set to rotate the polarization by $90^\circ$, which ensures equal path-length of the two spatial modes upon recombination at the second set of BD. The degree of loss for each polarization is then proportional to the offset of the corresponding HWP from this setting. Finally, a series of quarter-wave plates (QWP), HWP and polarizing beam splitter (PBS) is used to perform the Bell measurements. Note: additional polarizers may be introduced before the interferometer to produce high quality separable states.}
\label{fig:setup}
\end{figure}

Experimentally, we generate apparent super-quantum correlations based on the effect of polarization-dependent loss in a post-selected Bell-test experiment~\cite{Berry2010}, see Fig.~\ref{fig:setup}a). We use photon pairs created from a continuous-wave pumped spontaneous parametric down-conversion source in a polarization Sagnac design~\cite{Kim2006,Fedrizzi2007}, as illustrated in Fig.~\ref{fig:setup}b). Using this approach we obtain photon pairs with very high efficiency and in a continuously tunable fashion that enables us to produce any bipartite quantum state~\cite{Smith2012}.

In particular, we used the maximally entangled state $\ket{\psi^+}=\left(\ket H \ket V + \ket V \ket H \right)/\sqrt{2}$ as the initial state, where $\ket{H/V}$ represent horizontal and vertical polarization, respectively. For comparison, we also considered the corresponding fully decohered and thus separable state $\rho_\text{sep}=\left(\ketbra{HV}+\ketbra{VH}\right)/2$. This state was produced as a mixture of the two pure state components $\ket{HV}$ and $\ket{VH}$ by probabilistically mixing the respective coincidence counts.

The initial state is then subjected to polarization-dependent loss, introduced to the system by means of a Jamin-Lebedev polarization-interferometer, which allows individual control of the degree of loss for each polarization mode for both Alice and Bob, see Fig.~\ref{fig:setup}c). In the symmetric case considered here the loss was parametrized by a single parameter $\kappa$, where $\kappa{=}0$ corresponds to the loss-free scenario and $\kappa{=}1$ means complete loss of one polarization. With this setup we simulated correlations of increasing strength, ranging from classical to quantum and close to maximal non-signaling as discussed in detail in the methods section.

Using these correlations we investigated the information causality protocol on up to four levels (corresponding to a $16$-bit data-set for Alice) with $1$-bit of communication. Crucially, we implemented the protocol in Fig.~\ref{fig:ICMotivation}b) on a shot-by-shot basis, rather than estimating the performance from coincidence probabilities. For this we used an AIT-TTM8000 time-tagging module with a temporal resolution of $82$~ps to register the single photon counts for all possible outcomes. From this data, using passive feed-forward, i.e.\ at the processing stage, we were able to reconstruct over $10^5$ individual trials of the protocol for each of the $21$ settings of uniformly increasing $\kappa$.

\section{Results}
At a correlation strength of $S{=}3.874(5)$, the information available to Bob is at least $I{\geq}1.86(2)$ bits, despite only receiving $1$ bit from Alice. For four nesting levels of the protocol we establish lower bounds as high as $I{\geq}7.47(11)$ bit, which violates the information causality inequality $I\leq 1$ by almost $60$ standard deviations.
Similarly for weaker correlations, Bob has more information available than contained in Alice's message for all nesting levels as soon as the correlation strength surpasses $S \approx 3.5$. The fact that this value is significantly higher than Tsirelson's bound of $S_{\textsc{q}}\approx 3.41$ emphasizes that the quantity $I$ only recovers this bound in the asymptotic limit $n\to\infty$.

\begin{figure}[h!]
\begin{center}
\includegraphics[width=1\columnwidth]{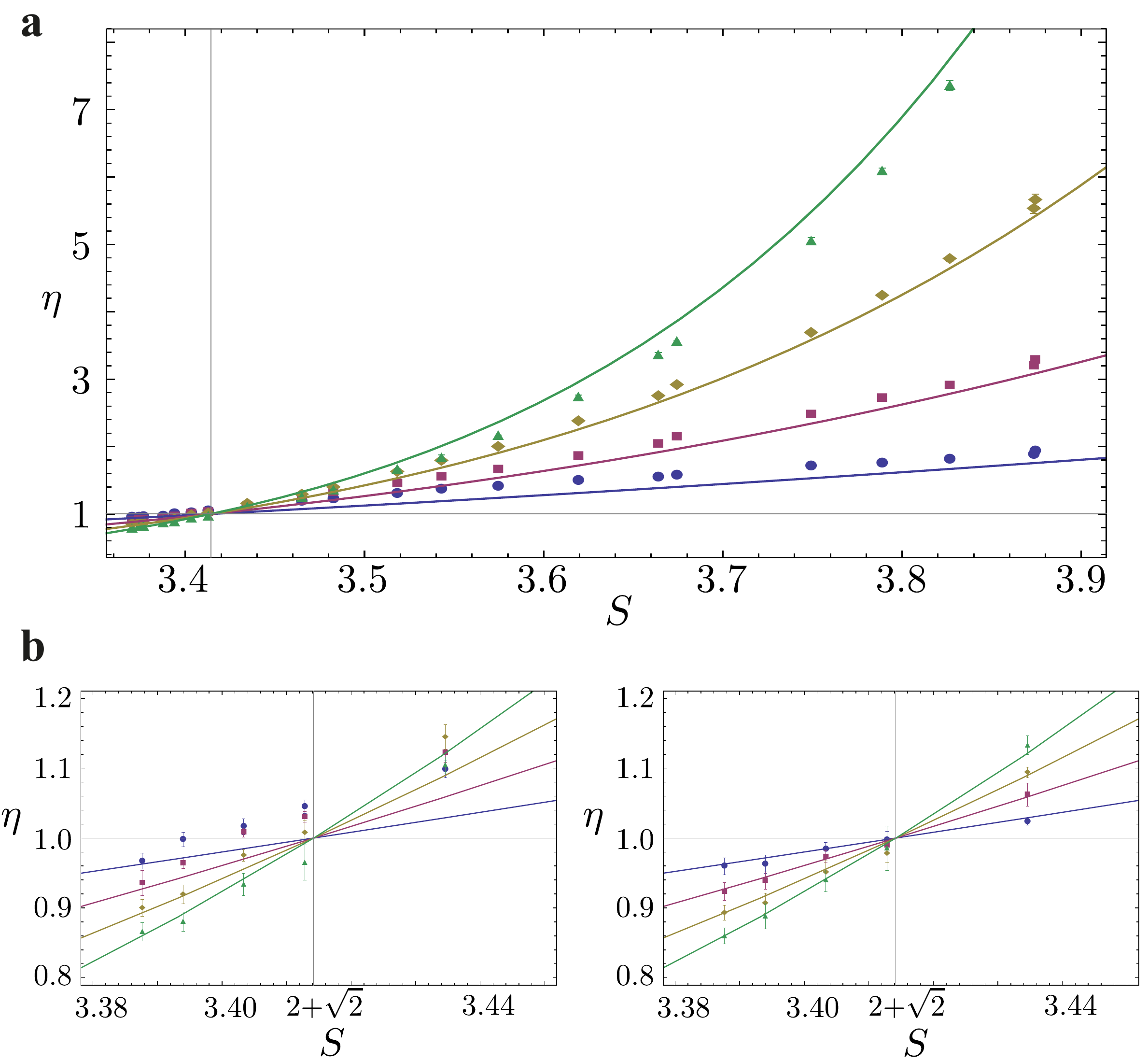}
\end{center}
\caption{\textbf{Experimental results for the efficiency in the information causality protocol.} \textbf{a)} Shown is the efficiency of the protocol for increasing strength of correlation, see methods section. The data points represent $n=1$ (blue circles), $n=2$ (red squares), $n=3$ (yellow diamonds) and $n=4$ (green triangles) levels in the protocol, where at each level a random dataset $\{a_i\}$ was used. Error-bars represent the standard deviation of 5 individual runs of every protocol. The lines correspond to theoretical expectations for the given correlation strength. \textbf{b)} A zoom into the region where our data violates Tsirelson's bound (indicated by the grey, vertical line). Our data violates the bound of $\eta{\leq}1$ already before the correlation strength surpasses Tsirelson's bound, which is a result of a finite sample size and the particular choice of random dataset, see Sec~\ref{Sec:Supp1}. In the right panel, the same plot for isotropic correlations obtained from using the protocol of Ref.~\cite{Masanes2006} shows very good agreement with the theoretical predictions.}
\label{fig:resultsIC:RAC}
\end{figure}

IIn the following we therefore consider an alternative figure of merit, motivated by identifying the protocol in Fig.~\ref{fig:ICMotivation}b) as a special case of a so-called random access code~\cite{Al-Safi2011}. Using similar ideas as in Ref.~\cite{Pawlowski2009}, the efficiency of this task can be bounded by
\begin{equation}
\eta=\sum_{k=0}^{N-1} (2P_k -1)^2 \leq 1 ,
\label{eq:RACbound}
\end{equation}
which thus also encompasses the principle of information causality~\cite{Al-Safi2011}. This bound, however, can indeed be saturated by quantum states for any size of Alice's dataset, as illustrated in Fig.~\ref{fig:resultsIC:RAC}. Note that our data violates the bound before the correlations surpass Tsirelson's bound. This is a result of a slight anisotropy in the simulated correlations due to experimental imperfections and a resulting bias for certain data-sets. It is not present when considering isotropic correlations, see Fig.~\ref{fig:resultsIC:RAC}b). Crucially, this highlights the dependence of both figures of merit~\eqref{eq:ICbound} and~\eqref{eq:RACbound} to the specific random choice of Alice's data-set.

In particular, the separable state used in the simulation produces entanglement-like correlations for one measurement choice of Alice and uncorrelated outputs for the other, see Fig.~\ref{fig:Correlations}. Hence, depending on the choice of data-set the figures of merit $\eta$ and $I$ might resemble the behavior expected for an entangled state, for a completely mixed state or, for higher nesting level, anything in-between. Only when averaging over all possible datasets, $\{a_i\}$, for a given level or employing the ``depolarization'' protocol introduced in Ref.~\cite{Masanes2006} to make the correlations isotropic without changing the CHSH value, can the quantities~\eqref{eq:ICbound} and~\eqref{eq:RACbound} be used as reliable figures of merit, see Fig.~\ref{fig:Correlations} and~\ref{fig:ResultsDepolSep-I}. Note, however, that anisotropic super-quantum correlations (averaged over all datasets) do not necessary violate Tsirelson's bound. In this case the principle of information causality cannot be probed using the depolarization approach, since it would result in isotropic correlations and information causality would not be violated.

\section{Discussion}
In contrast to the full set of no-signaling correlations, and the set of classical correlations, which both have the form of a well-characterized polytope, much less is known about the quantum set~\cite{Barrett2005,Brunner2014}. 
Understanding the set of quantum correlations theoretically and characterizing it experimentally should thus be a primary aim from a practical as well as a fundamental perspective. Information causality, which has been proposed as a physical principle to reconstruct the set of quantum correlations, has already proven successful in recovering the famous Tsirelson bound. This limit of quantum correlations, however, is only one extremal point on the continuous boundary and there exist correlations below it, which nevertheless do not admit a quantum description~\cite{Allcock2009}. Information causality also rules out such correlations for some 2-dimensional slices of the full (8-dimensional) no-signaling polytope, while it does not for other slices~\cite{Allcock2009}. This shortcoming, nevertheless is not definite and might just be a result of a suboptimal protocol in Fig.~\ref{fig:ICMotivation}b).

A violation of information causality would in particular imply that the tested theory does not admit a suitable measure of one of the most elementary information theoretic quantities: entropy~\cite{Al-Safi2011,Dahlsten2012}. Such a measure is assumed to be consistent with the classical limit and such that the entropy change $\Delta H$ of a composite system $XY$ satisfies $\Delta H(XY)\geq \Delta H(X) + \Delta H(Y)$ under local evolution of the subsystems $X$ and $Y$. Hence, a failure of these requirements could be interpreted as allowing for the generation of non-local correlations via local transformations.
Similar consequences might also arise from the violation of alternatives to information causality, which are more or less successful in recovering part of the quantum boundary. Examples include the principles of local orthogonality~\cite{Fritz2013}, the requirement that the theory has a suitable classical limit~\cite{Navascues2009} or that certain communication~\cite{VanDam2005,Brassard2006,Brunner2009} or computational tasks~\cite{Linden2007} are non-trivial.

Our method of simulating super-quantum correlations could be adapted to explore some of these alternative principles as well. Of particular interest, however, would be a test of information causality in the multipartite case, since most of the above principles are formulated in the bipartite setting, which is bound to fail in recovering the full quantum boundary due to the existence of multipartite super-quantum correlations, which obey every bipartite principle~\cite{Gallego2011,Yang2012}. While there are studies of information causality for higher-dimensional systems, which strengthen its position as a physical principle that determines quantum correlations~\cite{Cavalcanti2010}, a suitable generalization to the multipartite case is still an open problem.

As highlighted by our experiment, special focus has to be put on anisotropic regions of the no-signaling polytope. Specifically we find that the introduced figures of merit are not valid in a single instance of the protocol and have to be averaged over all possible datasets or estimated from the depolarized, isotropic data. This subtle, but very important detail is clearly highlighted by our experimental results, where we show how even a small amount of imbalance can result in a violation of the principle by quantum states for a specific choice of parameters, while obeying the principle on average.

\section{Methods}
\begin{figure}[h!]
\begin{center}
\includegraphics[width=1\columnwidth]{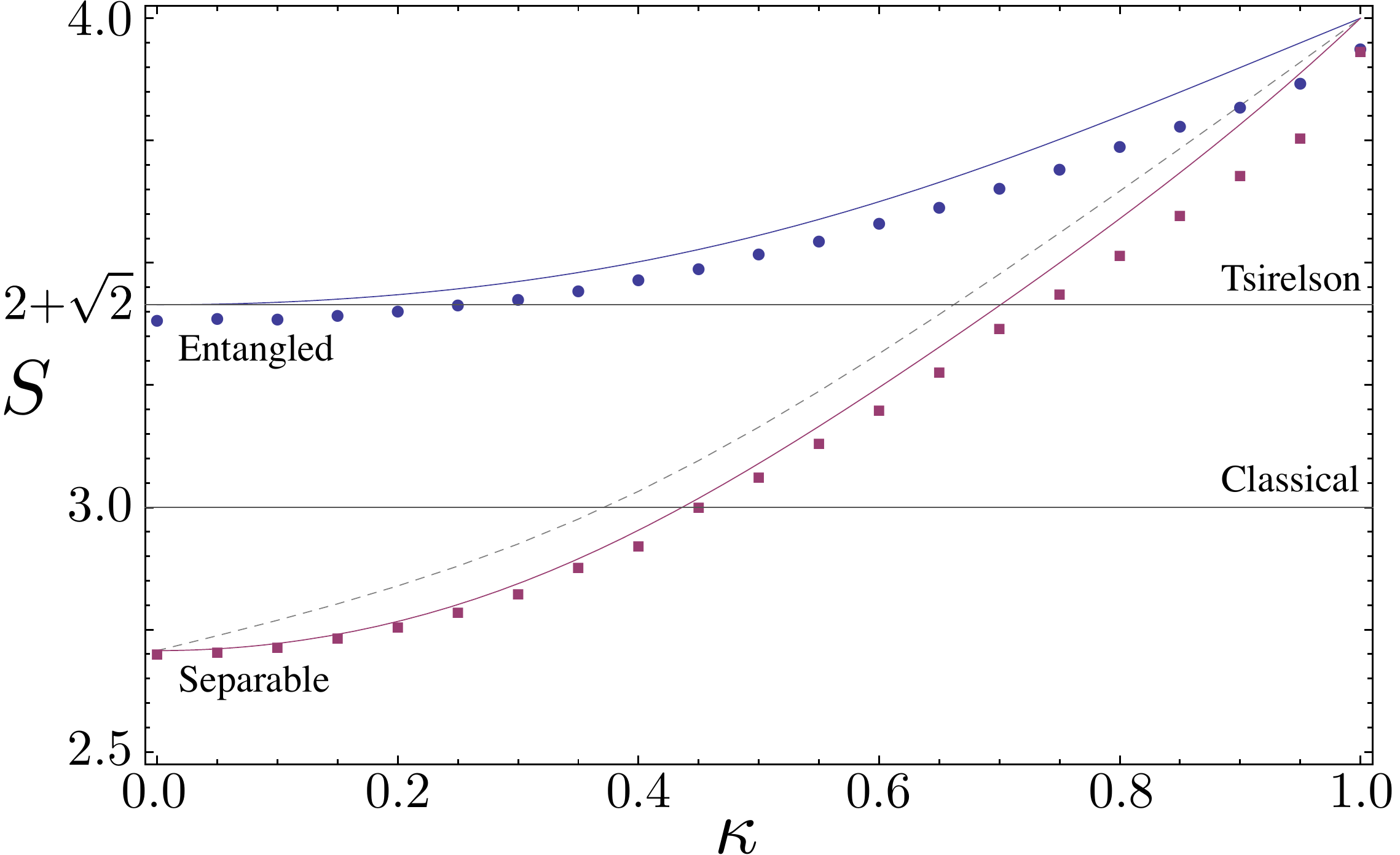}
\end{center}
\caption{\textbf{Experimental results.} Shown are the experimentally obtained values for the CHSH-parameter $S$ for both the entangled state $\ket{\psi^+}$ (blue circles) and the separable state $\rho_\text{sep}$ (red squares), together with the theoretical predictions (blue and red lines, respectively) for these states, versus the amount of polarization-dependent loss as parametrized by $\kappa$. The gray dashed line represents the theoretical expectation for the optimal separable state for a given amount of loss. In the experiment we observe a violation of Tsirelson's bound for $\kappa\ge 0.3$. Interestingly, we identify a region ($0.3{\leq}\kappa{\leq}0.372$) where the quantum bound of the inequality is violated, while the classical bound still holds for all separable states. With the chosen, fixed, separable state $\rho_\text{sep}$ we observe a first violation at $\kappa=0.5$. Errors from a Monte-Carlo sampling of the Poissonian counting statistics are not visible on the scale of this plot.}
\label{fig:results}
\end{figure}

Examining the results of a CHSH-inequality test make it clear where our data crosses the boundary of the quantum set. In our investigation we focused on the scenario of a fixed maximally entangled state $\ket{\psi_+}$ in situations with different amounts of loss, as shown in Fig.~\ref{fig:results}. We further considered the state $\rho_{\text{sep}}$, which resembles the state $\ket{\psi_+}$ after full decoherence as might happen during propagation between Alice and Bob. This allows for an intuitive comparison between the entangled and unentangled case.

The tested inequality has the form of a CHSH-inequality with measurements in the \textsc{yz}-plane of the Bloch sphere. For the lossless case $\kappa=0$, Alice's and Bob's measurements can be viewed as the application of appropriate basis-rotations (around the \textsc{x}-axis) followed by projective measurements in the $\ket{H/V}$-basis. These rotations can also be seen as phase-gates in the diagonal polarization basis $\ket{\pm}=\frac{1}{\sqrt{2}}(\ket{H}\pm\ket{V})$. In the case where polarization dependent loss is present, these phase-gates become non-unitary. They act as the identity on the state $\ket{u}=\left(\sqrt{1+\kappa}\, \ket{H}{+}\sqrt{1-\kappa}\, \ket{V}\right)/\sqrt{2}$ and impose a phase on the non-orthogonal state $\ket{w}=\left(\sqrt{1+\kappa}\, \ket{H}{-}\sqrt{1-\kappa}\, \ket{V}\right)/\sqrt{2}$, where $\kappa=\braket{u}{w}$. The precise relation between $\kappa$ and the degree of loss is discussed in Ref.~\cite{Berry2010}.
As non-unitary operations can only be performed non-deterministically, postselection on success is required, which results in the observation of apparent super-quantum correlations. Finally, we use the first step of the depolarization procedure in Ref.~\cite{Masanes2006} to symmetrize the simulated correlations, while preserving their possible anisotropy.

Curiously, we note that moderate polarization-dependent loss can lead to super-quantum correlations for entangled states without invalidating the CHSH inequality for separable states, as suggested in Ref.~\cite{Berry2010}. This observation even holds when optimizing the separable state for maximal CHSH-value, for each degree of loss~\cite{Berry2010}. Note, however, that these results were obtained using the same measurements for both separable and entangled states, whereas arbitrary hidden variable theories would allow arbitrary measurements.

Figure~\ref{fig:results} illustrates the obtained values of the CHSH parameter $S$ and compares them to the ideal case, which, for the initially entangled state, is described by
\begin{equation}
S_{\ket{\psi_+}}(\kappa) = 3 \frac{\kappa^2-\cos{\frac\Theta 2}}{2(1-\kappa^2 \cos{\frac\Theta 2})} - \frac{\kappa^2-\cos{\frac {3\Theta}{2}}}{2(1-\kappa^2 \cos{\frac{3\Theta}{2}})} +2.
\end{equation}
Here $\Theta$ is a function of $\kappa$, which can be analytically approximated by $\Theta{=}\pi(17{+}\cos(\pi\kappa))/12$, as discussed in Ref.~\cite{Berry2010}.

We experimentally violate Tsirelson's bound by more than $7$ standard deviations, $S{=}3.423(1)$ at a loss parameter of $\kappa{=}0.3$. At this point, the achieved value for the unentangled state, $S{=}2.821(2)$, is indeed well below the classical bound of $3$ and even the optimal unentangled state does not violate the inequality until $\kappa\approx 0.37$. In the region $0.3\leq \kappa\leq 0.37$ it is therefore possible to exploit super-quantum correlations from entangled states while unentangled states still appear classical. With increasing loss, both states eventually violate Tsirelson's bound and approach the numerical maximum of $S=4$, with experimental values of $S=3.9341(6)$ and $S=3.929(2)$ for the entangled and separable state, respectively. The increasing deviation from the theoretical predictions in Fig.~\ref{fig:results} is a result of the decreasing signal-to-noise ratio in the single-photon detectors for high-loss settings.

Related experiments have observed apparent violations of Tsirelson's bound as a consequence of explicit violations of the detection loophole~\cite{Tasca2009} or the fair-sampling assumption~\cite{Romero2013}. The latter is in fact typically violated if the quantum system of interest has more (possibly) correlated degrees of freedom than those tested in the Bell-inequality~\cite{Dada2011}. Violation of Tsirelson's bound has also been considered as an intermediate step in deriving three-qubit inequalities~\cite{Cabello2002}.

\subsection*{Acknowledgments}
We thank Tim Ralph and Cyril Branciard for helpful discussions. We also thank the team from the Austrian Institute of Technology for kindly providing the time-tagging modules for this experiment. We acknowledge financial support from the ARC Centres of Excellence for Engineered Quantum Systems (CE110001013) and Quantum Computation and Communication Technology (CE110001027). D.W.B.\ is funded by an Australian Research Council Future Fellowship (FT100100761), and A.F by an Australian Research Council Discovery Early Career Researcher Award (DE130100240). A.G.W.\ acknowledges support from a UQ Vice-Chancellor's Senior Research Fellowship.

\onecolumngrid
\clearpage
\renewcommand{\theequation}{S\arabic{equation}}
\renewcommand{\thefigure}{S\arabic{figure}}
\renewcommand{\thetable}{\Roman{table}}
\renewcommand{\thesection}{S\Roman{section}}
\setcounter{equation}{0}
\setcounter{figure}{0}
\setcounter{section}{0}
\begin{center}
{\bf \large Supplemental Material}
\end{center}
\twocolumngrid
Here we discuss in more detail the subtleties of the information causality protocol for individual choices of the data-set for Alice. In particular we will discuss the effect of anisotropic correlations of decohered entangled states and present results for the case where these are transformed to isotropic correlations by incorporating the protocol proposed in Ref.~\cite{Masanes2006} into our experiment.

\section{Separable state correlations and the choice of data-set}
\label{Sec:Supp1}
As discussed previously, the experiment was performed for both entangled and separable initial states. While the entangled state allows for generation of super-quantum correlations in the range $3.379(1) \leq S \leq 3.9341(6)$, the separable state covers a larger range of $2.698(2)\leq S\leq 3.929(2)$. This range in particular includes a large part of the set of classical ($S\leq 3$) and quantum correlations ($S\leq 3.41$). Here it is important to reiterate the form of the used state. Since it has been chosen as the fully decohered version of $\ket{\psi_+}$ it still retains correlations of the same strength in one basis, while correlations in any orthogonal basis are lost---a classically correlated state.

\begin{figure}[h!]
\begin{center}
\includegraphics[width=1\columnwidth]{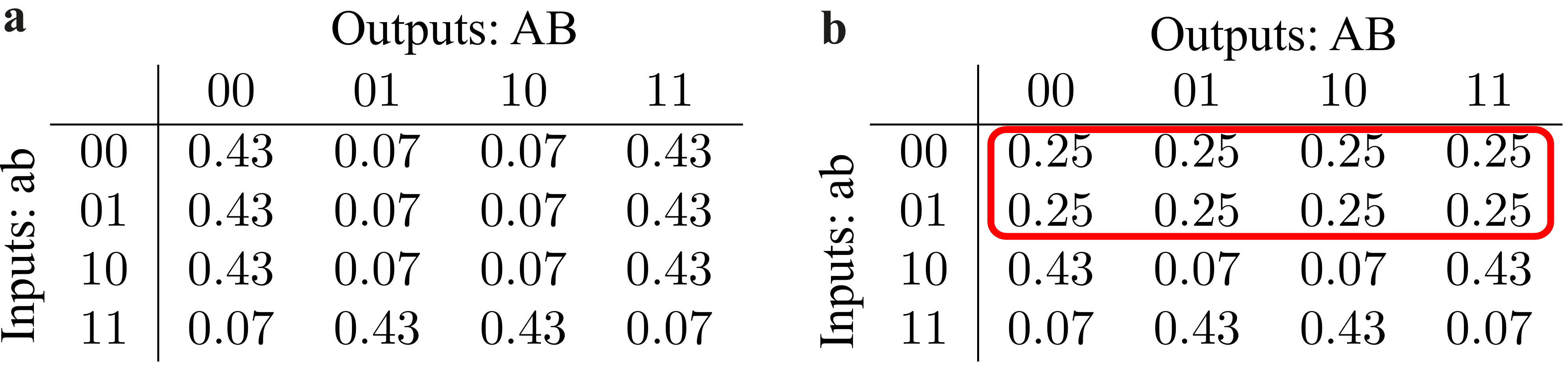}
\end{center}
\caption{\textbf{Measurement probabilities for the lossless case $\kappa=0$.} Shown are the theoretically expected measurement probabilities for both the \textbf{a)} entangled and \textbf{b)} separable state for a lossless Bell-test experiment. Note that the correlations are the same for the two states when Alice chooses to measure $1$, but there are no correlations for the separable state when she measures $0$.}
\label{fig:Correlations}
\end{figure}

An important consequence of this feature, as discussed in the main text, is that the success probability in the information causality protocol then depends on the specific choice of data-set for Alice. As an example consider the simplest instance, where Alice has two bits $a_0$ and $a_1$. Alice uses $a_0\oplus a_1$ as her input and Bob uses $b\in\{0,1\}$. The probability of success in this scenario will be the same as for the entangled state whenever $a_0\oplus a_1=1$ (that is for the data-sets $\{0,1\}$ and $\{1,0\}$) and the same as for random guessing ($1/2$) in the other two cases, where $a_0\oplus a_1=0$. Although Ref.~\cite{Pawlowski2009} discussed the related case where the probability of success may depend on Bob's choice, the feature revealed here has important practical implications, since the calculation of the figures of merit always requires to consider all of Bob's possible choices, while only one data-set for Alice has to be considered. Depending on the choice of data-set, any degree of efficiency can be achieved with classical states.

There are several ways to circumvent these problems. Clearly, averaging over all possible data-sets will recover the performance expected from the CHSH-value corresponding to the respective state. However, as the possible choices for Alice's data-set grow doubly-exponentially, this approach is typically unfeasible, in particular when testing the protocol on a shot-by-shot basis. 
Here we employed the ``depolarization'' protocol introduced by Ref.~\cite{Masanes2006}, which takes any set of correlations to an isotropic one without changing the CHSH-value, using 3 bits of shared randomness. The drawback of this method, however, is that it precludes a test of information causality in the anisotropic regime of super-quantum correlations below Tsirelson's bound, see Ref.~\cite{Allcock2009}.

\section{Results for isotropic correlations}
\label{Sec:Supp2}
\begin{figure}[h!]
\begin{center}
\includegraphics[width=1\columnwidth]{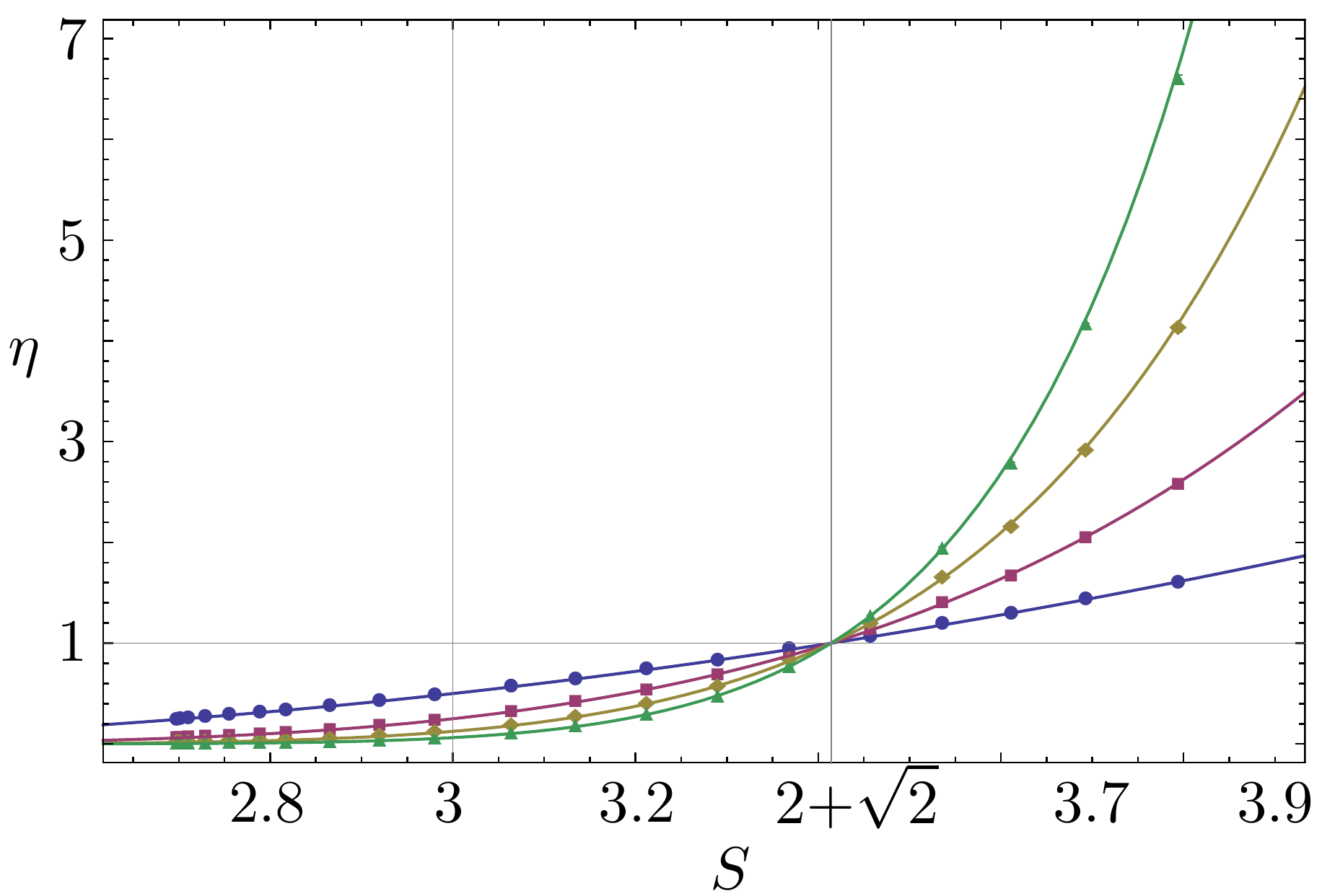}
\end{center}
\caption{\textbf{Experimental results for the mutual information gain in the information causality protocol.} Shown is the lower bound on the mutual information gain in the protocol for increasing strength of isotropic correlation. The data points represent $n=1$ (blue circles), $n=2$ (red squares), $n=3$ (yellow diamonds) and $n=4$ (green triangles) levels in the protocol, where at each level a random dataset $\{a_i\}$ was used. Error-bars represent the standard deviation of 5 individual runs of every protocol. The lines correspond to theoretical expectations for the given correlation strength.}
\label{fig:ResultsDepolSep-I}
\end{figure}

\begin{figure}[h!]
\begin{center}
\includegraphics[width=1\columnwidth]{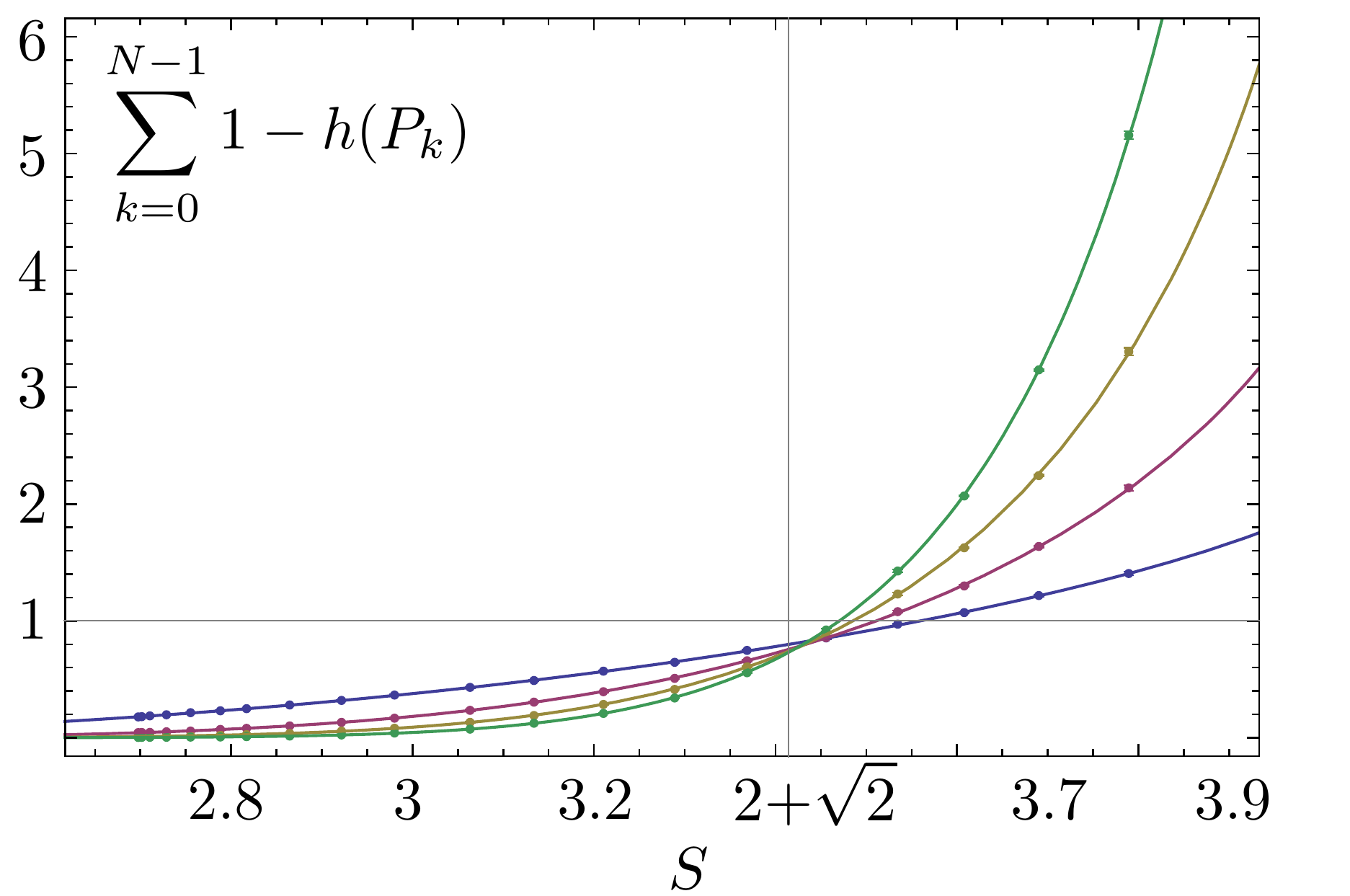}
\end{center}
\caption{\textbf{Experimental results for the efficiency in the information causality protocol.} Shown is the efficiency of the protocol for increasing strength of isotropic correlation. The data points represent $n=1$ (blue circles), $n=2$ (red squares), $n=3$ (yellow diamonds) and $n=4$ (green triangles) levels in the protocol, where at each level a random dataset $\{a_i\}$ was used. Error-bars represent the standard deviation of 5 individual runs of every protocol. The lines correspond to theoretical expectations for the given correlation strength.}
\label{fig:ResultsDepolSep-RAC}
\end{figure}

Figure~\ref{fig:ResultsDepolSep-I} shows the experimental lower bounds on the mutual information measure $I$ for the isotropic correlations obtained from the depolarization protocol applied to the initially separable state. Figure~\ref{fig:ResultsDepolSep-RAC} shows the random access code efficiency $\eta$ for the same data. In both cases we observe excellent agreement with the theoretical predictions.

\end{document}